\documentclass{article}
\usepackage{spconf,amsmath,graphicx}
\usepackage{xcolor,array}

\newcommand\blfootnote[1]{%
	\begingroup
	\renewcommand\thefootnote{}\footnote{#1}%
	\addtocounter{footnote}{-1}%
	\endgroup
}

\title{AEC in a NetShell: \\On Target and Topology Choices \\ for FCRN Acoustic Echo Cancellation}
\name{Jan Franzen, Ernst Seidel, Tim Fingscheidt}
\address{Institute for Communications Technology, Technische Universit\"at Braunschweig,\\ Schleinitzstr. 22, 38106 Braunschweig, Germany\\
	{\small \tt \{j.franzen, ernst.seidel, t.fingscheidt\}@tu-bs.de}}

\begin{document}
\ninept
\maketitle
\begin{abstract}
\vspace{-2mm}
Acoustic echo cancellation (AEC) algorithms have a long-term steady role in signal processing, with approaches improving the performance of applications such as automotive hands-free systems, smart home and loudspeaker devices, or web conference systems.
Just recently, very first deep neural network (DNN)-based approaches were proposed with a DNN for joint AEC and residual echo suppression (RES)/noise reduction, showing significant improvements in terms of echo suppression performance. 
Noise reduction algorithms, on the other hand, have enjoyed already a lot of attention with regard to DNN approaches, with the fully convolutional recurrent network (FCRN) architecture being among state of the art topologies.
The recently published impressive echo cancellation performance of joint AEC/RES DNNs, however, so far came along with an undeniable impairment of speech quality. 
In this work we will heal this issue and significantly improve the near-end speech component quality over existing approaches. Also, we propose for the first time---to the best of our knowledge---a pure DNN AEC in the form of an echo estimator, that is based on a competitive FCRN structure and delivers a quality useful for practical applications.
\end{abstract}
\vspace{-1.6mm}
\begin{keywords}
acoustic echo cancellation, echo suppression, convolutional neural network, ConvLSTM
\blfootnote{\scriptsize$\copyright$ 2021 IEEE. Personal use of this material is permitted. Permission from IEEE must be obtained for all other uses, in any current or future media, including reprinting/republishing this material for advertising or promotional purposes, creating new collective works, for resale or redistribution to servers or lists, or reuse of any copyrighted component of this work in other works.}
\end{keywords}

\vspace{-4mm}

\section{Introduction}
\label{sec:intro}
\vspace{-2mm}

Applications such as automotive hands-free systems, smart home and loudspeaker devices, web conference systems, and many more share a similar underlying challenge: The microphone signal picks up an undesired echo component stemming from the system's own loudspeakers. With acoustic echo cancellation (AEC) algorithms having a steady role in signal processing over the past decades, these algorithms typically deploy an adaptive filter to estimate the impulse response (IR) of the loudspeaker-enclosure-microphone~(LEM) system. The echo component is then estimated and subsequently subtracted from the microphone signal to obtain a widely echo-free enhanced near-end speech signal.

Traditional AEC algorithms~\cite{haensler_acousticechocontrol, Lee_blockbased-filters, shin_NLMS-AP-algos} have a long-term role in signal processing, with the approaches steadily evolving, resulting in renowned algorithms using normalized least mean squares (NLMS) algorithm~\cite{steinert_lowdelayhandsfree} or the Kalman filter~\cite{enzner_vary_fdaf, franzen_LowDelayICC_INTERSPEECH}, and including residual echo suppression (RES) approaches~\cite{KuechEnzner_StateSpacePartitionedFDAF, franzen_RES_ICASSP}.
In the recent past, neural networks---especially convolutional neural networks---have shown significant performance in speech enhancement in general, e.g., the work of Strake et al.~\cite{strake_SingleStage_ICASSP} for noise reduction. However, AEC has only seen very few data-driven approaches so far. Initially, only networks for RES were among them~\cite{Schwarz_NN_FF_RES, carbajal_RES_ICASSP}. 

Just recently, a fully learned AEC was proposed by Zhang et~al.~\cite{wang_NN_AEC_18, wang_NN_AEC_19}, revealing an impressive echo cancellation performance. An interesting aspect of these works is the way the AEC problem is tackled. It is treated as source separation approach with the networks being trained to directly output the estimated enhanced signal.

The difficulty with AEC DNNs is, however, that they so far come along with an undeniable impairment of the near-end speech component quality. In this work, we will investigate this issue with a set of experiments to show and reveal trade-offs between different performance aspects in terms of echo suppression, noise reduction, and near-end speech quality. 
With the fully convolutional recurrent network (FCRN)~\cite{strake_SingleStage_ICASSP, zhao_CNN} and its proven capability of autoencoding speech at high fidelity as a basis, we will introduce several DNN AEC architectures overcoming earlier problems, thereby significantly improving over existing approaches. We will provide useful insights into network design choices, giving the reader guidance on the not yet widely explored field of DNN AEC.

The remainder of this paper is structured as follows: In Section~\ref{sec:setup}, a system overview including the framework and general network topology is given. The training and different experimental variants including novel network topology choices are described in Section~\ref{sec:variants}.
In Section~\ref{sec:results}, the experimental validation and discussion of all approaches is given. Section~\ref{sec:conc} provides conclusions.

\vspace{-3.5mm}
\section{Network Topology, \\Simulation Framework, and Data}
\label{sec:setup}
\vspace{-1.7mm}

\begin{figure*}[htb]
	\centering
	\includegraphics[width=\textwidth]{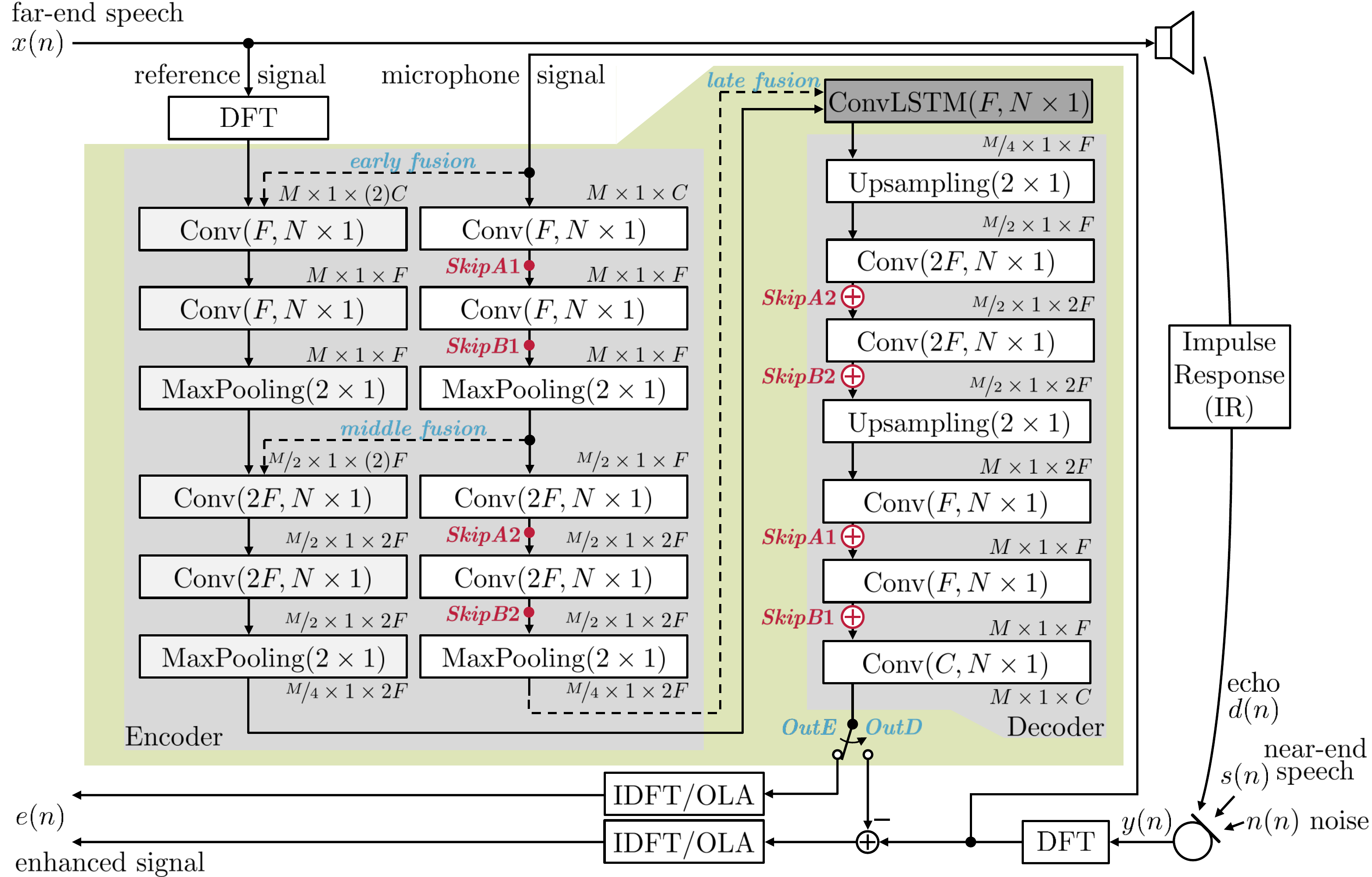}
	\vspace{-7mm}
	\caption{System model and network with various options for skip connections, encoder fusion and training targets. The parameters of the convolutional layers are Conv(\textit{\# filters, kernel dimensions}), maximum pooling MaxPooling(\textit{pool dimensions}) over the feature axis, and likewise for upsampling. Skip connection variants (\textit{\textbf{none, SkipA, SkipB}}) are indicated by the respective start and end positions with the same naming. Only a single dashed path (-\,-\,-) is involved at a time, determining either \textit{\textbf{early}}, \textit{\textbf{middle}}, or \textit{\textbf{late fusion}} of the reference and microphone signals in the encoder. Training target options are indicated by the switch positions \textit{\textbf{OutE}} (speech target) or \textit{\textbf{OutD}} (echo target).}
	\label{fig:network}
	\vspace{-5mm}
\end{figure*}

\subsection{Novel FCRN Network Topology}
\label{sec:network}
\vspace{-1mm}
In contrast to traditional adaptive filters, the AEC itself is performed by a neural network in this work. Basis for our experiments is the well-performing fully convolutional recurrent network (FCRN) encoder-decoder structure proposed for noise reduction in~\cite{strake_SingleStage_ICASSP}. However, we introduce important AEC specifics into the network topology. Our proposed network is depicted within the green box of Figure~\ref{fig:network}, operates on discrete Fourier transform (DFT) inputs $X_\ell(k)$ with frame index $\ell$ and frequency bin $k$, and contains a couple of novelties: Originally consisting of only one encoder (i.e., here, most likely comparable to performing early fusion with the microphone signal DFT $Y_\ell(k)$ and following only the respective dashed signal path), we investigate a parallel second encoder (portion) consisting of up to two times two convolutional layers, followed by maximum pooling over the feature dimension with stride~$2$. The first two convolutional layers use $F$~filter kernels of size $N\!\times\!1$ (convolution over the feature axis), whereas the latter two use $2F$~filter kernels of the same size. Leaky ReLU activations \cite{leakyReLU} are used for these layers. For easier readability, feature dimensions can be seen at the in- and output of each layer, denoted as \textit{feature axis $\times$ time axis $\times$ number of feature maps}. During inference, the network is subsequently processing single input frames, which is indicated by the time axis value being set to~$1$. 

At the bottleneck, right after the encoder, where the feature axis reaches its maximum compression of $M/4$, a convolutional LSTM~\cite{ConvLSTM} with $F$~filter kernels of size $N\!\times\!1$ is placed, enabling the network to model temporal context. The decoder is set up exactly as inverse to the encoder and followed by a final convolutional layer with linear activations to yield the final output of dimension $M \times 1 \times C$.
To extract input features and training targets for the structures given in Figure~\ref{fig:network}, at a sampling rate of $f_s \! = \! 16$\,kHz a frame length of $K \! = \! 512$~samples is used and the frame shift is set to $256$~samples. By applying a square root Hann window and $512$-point DFT, complex spectra are obtained. Separated into real and imaginary parts, and zero-padded to feature maps of height $M\,=\,260$, this leads to $C\,=\,2$ channels for the reference, the microphone, and the estimated echo or (clean) speech signal. 

\vspace{-3.8mm}
\subsection{Simulation Framework and Data}
\label{sec:data}
\vspace{-1.2mm}

To model the acoustic setup shown in Figure~\ref{fig:network}, we adopt the procedure described in~\cite{wang_NN_AEC_19} with some modifications. 
Thus, to model typical single- and double-talk scenarios, far-end speech~$x(n)$ and near-end speech~$s(n)$ are set up using the TIMIT dataset~\cite{TIMIT}.
Background noises~$n(n)$ are taken from the QUT dataset~\cite{QUT} for training and validation, while babble, white noise, and the operations room noises are used from the NOISEX-92 dataset~\cite{noisex} for the test set. Noise~$n(n)$ is superimposed with near-end speech~$s(n)$ at the microphone, and echo signals~$d(n)$ are generated by imposing  
loudspeaker nonlinearities~\cite{wang_NN_AEC_19} on far-end signals~$x(n)$ and convolving them with impulse responses (IRs) of $512$~samples length. The IRs are created using the image method~\cite{image_method} with reverberation times $T_{60} \! \in \! \{0.2, 0.3, 0.4\}$\,s for training and validation, and $0.2$\,s for test mixtures, thereby following~\cite{wang_NN_AEC_19}. A test with additional real IRs is omitted here for space reasons,
since it was impressively shown in~\cite{wang_NN_AEC_19} that apparently for DNN AECs comparable results are obtained for both, real and simulated IRs.
For a broad variety of simulations, signal-to-echo ratios (SER) are selected randomly between $\{-6, -3, 0, 3, 6, \infty\}$\,dB per mixture and signal-to-noise ratios (SNR) are selected randomly between $\{8, 10, 12, 14, \infty\}$\,dB per mixture. Note, that we included~$\infty$~dB to the SER and SNR values, since for a practical application it is absolutely mandatory that the absence of echo or noise can be handled by the network as well. 
In our setup this leads to a total of 3000 training, 500 evaluation, and 280 test mixtures, whereas the latter---differing from~\cite{wang_NN_AEC_19}---consist of \emph{unseen} speakers from the CSTR VCTK database~\cite{VCTK} with unseen utterances, impulse responses, and noise sequences. 
SER and SNR for the test mixtures are set to $0$\,dB and $10$\,dB, respectively. For deeper insights into the network performance, we additionally evaluate the test files but consisting either of echo only, or of near-end noise or near-end speech only. 

\vspace{-1.4mm}
\section{Experimental Variants and Training}
\label{sec:variants}
\vspace{-1.4mm}

\subsection{Training Target Variants}
\label{sec:targets}

One major question we investigate is rather significant and concerns the choice of the training targets. 
Here, \cite{wang_NN_AEC_19} differs from the traditional concept for AEC, where an estimated echo $\hat{d}(n)$ is generated, which is then subtracted from the microphone signal to obtain an (ideally) echo-free enhanced signal $e(n)$. In~\cite{wang_NN_AEC_18, wang_NN_AEC_19}, however, the echo problem is tackled by a source separation approach trained to directly output the estimated enhanced signal $E_\ell(k)$, thereby enabling two meaningful possibilities for the regression training targets $\overline{E}_\ell(k)$ in the DFT domain: The complex-valued target can either be chosen as $\overline{E}_\ell(k) \! = \! S_\ell(k) \! + \! N_\ell(k)$ (i.e., only echo cancellation is performed by the network), or just as $\overline{E}_\ell(k) \! = \! S_\ell(k)$ (i.e., echo \textit{and} noise cancellation are performed). 
This leads to the question which of the mentioned targets are best suited and if there are any tradeoffs to be dealt with.

Indicated by the network output switch in Figure~\ref{fig:network}, we investigate the two different variants of training targets with an MSE loss 
$J_\ell \! = \! \frac{1}{K}\!\sum_{k \in K} \!
| \hat{E}_\ell(k) - \overline{E}_\ell(k) |^2$
in the frequency domain (switch position \textbf{\textit{OutE}}), with $\hat{E}_\ell(k)$ being the respective network outputs. 
As the third variant, the MSE loss  
	$J_\ell \! = \! \frac{1}{K}\!\sum_{k \in K} \!
	| \hat{D}_\ell(k) - \overline{D}_\ell(k) |^2$
is applied using the echo component training targets $\overline{D}_\ell(k) \! = \! D_\ell(k)$ directly with subsequent subtraction from the microphone signal (switch position \textbf{\textit{OutD}}).

\vspace{-1.4mm}
\subsection{Skip Connection Variants}
\label{sec:skip}
Throughout this work we will experiment with different positions for skip connections reaching from encoder to decoder. The original model has a skip connection placed between the red marked points \textit{SkipB1} and another one between the points \textit{SkipB2}~\cite{strake_SingleStage_ICASSP}. Hereinafter, this setup will be denoted as \textit{\textbf{SkipB}}. With the varying dimensions of the feature maps, a second possibility is given by placing the skip connections in a symmetric manner, i.e., one between the points \textit{SkipA1} and another one between the points \textit{SkipA2}. This setup will be denoted as \textit{\textbf{SkipA}}. The last variant is to use no skip connections at all, which will be denoted as \textit{\textbf{NoSkips}} (---).

\vspace{-1.4mm}
\subsection{Encoder Fusion Variants}
\label{sec:encoder}

A traditional AEC algorithm uses the reference signal~$x(n)$ as input and replicates the IR with an adaptive filter. The microphone signal~$y(n)$ (or rather a thereon based error signal) serves as control input for the adaptive filter.
In contrast, for the network of~\cite{wang_NN_AEC_19}, as for our network when a combined encoder is used (early fusion), the feature maps of reference and microphone signal are directly concatenated at the network input, 
denoted as \textit{\textbf{EarlyF}}.  

However, the original idea of using an encoder-decoder structure is to allow the network to preprocess and find a suitable representation of its input signals throughout the encoder, which can then be well-handled by its bottleneck layer. At this point it is important to notice that convolutional layers with $N\!\times\!1$~kernels along the frequency axis are not able to model delay, which we consider of crucial importance for handling the temporal shift between reference and microphone signal. 
Since our main processing unit in the bottleneck layer is a convolutional LSTM that can indeed model delay, we experiment with performing fusion of the reference and microphone signals at different positions in the encoder. This shall allow the network to process microphone and reference signals separately to a certain degree, before the respective feature maps are concatenated and processed together throughout the remaining network. 

Two further variants for the encoder fusion will be considered: The first one is the middle fusion, in the following denoted as \textit{\textbf{MidF}}, where only the respective dashed signal path is involved.
The second variant duplicates the entire encoder and performs feature map concatenation at the input to the convolutional LSTM. Here, only the last respective dashed signal path is used. The method is denoted as \textit{\textbf{LateF}} in the experiments.

If middle or late fusion is performed in combination with skip connections, the skip connections are branched off of the microphone signal path as indicated in Figure~\ref{fig:network}. We also considered placing their starting points in the reference signal path, but as can be expected this does not lead to any meaningful results. When early fusion is performed, the skip connections are branched off of the respective positions of the common encoder.

\vspace{-1.4mm}
\subsection{Training Parameters}

Networks are trained with the Adam optimizer~\cite{adam_optimizer} using its standard parameters. The batch size and sequence length are set to~$16$ and~$50$, respectively. With an initial learning rate of $5 \! \cdot \! 10^{-5}$, the learning rate is multiplied with a factor of $0.6$ if the loss did not improve for $3$ epochs. Training is stopped when the learning rate drops below $5 \! \cdot \! 10^{-6}$ or if the loss did not improve for $10$~epochs. The number of parameters varies with the encoder fusion position, resulting in~$5.2$\,M and $5.6$\,M~parameters for \textit{\textbf{EarlyF}} and \textit{\textbf{MidF}}, respectively, and $7.1$\,M~parameters for \textit{\textbf{LateF}}.

\vspace{-1.4mm}
\section{Results and Discussion}
\label{sec:results}

\begin{table}[htbp]
	\caption{Experiment results: ERLE and deltaSNR given in~[dB], and PESQ MOS LQO for all models with \textbf{clean speech} training target \textit{\textbf{OutE}}: $\overline{E}_\ell(k) \! = \! S_\ell(k)$ following~\cite{wang_NN_AEC_19}. For deeper insights, the three columns on the right show the respective performance when only one component is present at the microphone. Best result per  measure is marked in \textbf{bold} font, second best is \underline{underlined}.}
	\label{tab:unseen_res_s_0}
	\vspace{2mm}
	\setlength\tabcolsep{0pt} 
	\begin{tabular}{p{14mm}<{\centering} | p{9mm}<{\centering} p{12mm}<{\centering} p{12mm}<{\centering} p{12mm}<{\centering} | p{9mm}<{\centering} | p{9mm}<{\centering} | p{8mm}<{\centering}}
		\textbf{Model/} & \multicolumn{4}{c|}{full mixture} & $d(n)$ & $n(n)$ & $s(n)$ \\
		\textbf{Skip} & \textbf{PESQ} & \textbf{ERLE}$_{\text{BB}}$ & \textbf{dSNR}$_{\text{BB}}$ & \textbf{PESQ}$_{\text{BB}}$ & \textbf{ERLE} & \textbf{dSNR} & \textbf{PESQ} \\
		\hline
		Kalman & 1.15 & 3.49 & -0.94 & \textbf{4.64} & 18.36 & --- & \textbf{4.64} \\
		\hline
		EarlyF/B & 1.52 & 19.49 & 11.05 & 2.82 & 11.83 & 32.94 & 3.24 \\
		EarlyF/A & 1.44 & 20.73 & 10.94 & 2.56 & 13.20 & \underline{33.33} & \underline{3.65} \\
		\textbf{EarlyF/---} & \textbf{1.57} & \textbf{25.85} & \underline{11.47} & 2.64 & \textbf{21.33} & 32.94 & 3.44 \\
		MidF/B & 1.52 & 20.87 & 10.24 & \underline{2.83} & 15.79 & 29.65 & 3.64 \\
		MidF/A & 1.49 & 21.00 & 11.25 & 2.70 & 15.45 & 27.12 & 3.32 \\
		MidF/--- & \underline{1.56} & 24.63 & 11.05 & 2.81 & 18.25 & 32.97 & 3.33 \\
		LateF/B & 1.45 & 23.27 & \textbf{11.68} & 2.62 & 17.77 & 27.87 & 3.44 \\
		LateF/A & 1.52 & 20.40 & 10.43 & 2.70 & 15.23 & 27.38 & 3.42 \\
		LateF/--- & 1.53 & \underline{24.80} & 10.81 & 2.66 & \underline{19.23} & \textbf{33.62} & 3.28 \\
	\end{tabular}
	
	\vspace{2mm}
	\caption{Experiment results for all models as in Table~\ref{tab:unseen_res_s_0}, but with \textbf{noisy speech} training target \textit{\textbf{OutE}}: $\overline{E}_\ell(k) \! = \! S_\ell(k)  + \! N_\ell(k)$. Best result per measure is marked in \textbf{bold} font, second best is \underline{underlined}. Additional result for best model EarlyF/A with separate subsequent noise reduction from~\cite{strake_SingleStage_ICASSP}, retrained on this work's data (EarlyF/A+).}
	\label{tab:unseen_res_sn_0}
	\vspace{2mm}
	\setlength\tabcolsep{0pt} 
	\begin{tabular}{p{14mm}<{\centering} | p{9mm}<{\centering} p{12mm}<{\centering} p{12mm}<{\centering} p{12mm}<{\centering} | p{9mm}<{\centering} | p{9mm}<{\centering} | p{8mm}<{\centering}}
		\textbf{Model/} & \multicolumn{4}{c|}{full mixture} & $d(n)$ & $n(n)$ & $s(n)$ \\
		\textbf{Skip} & \textbf{PESQ} & \textbf{ERLE}$_{\text{BB}}$ & \textbf{dSNR}$_{\text{BB}}$ & \textbf{PESQ}$_{\text{BB}}$ & \textbf{ERLE} & \textbf{dSNR} & \textbf{PESQ} \\
		\hline
		Kalman & 1.15 & 3.49 & -0.94 & \textbf{4.64} & 18.36 & --- & \textbf{4.64} \\
		\hline
		EarlyF/B & \textbf{1.51} & 8.99 & 0.63 & \underline{3.10} & 6.95 & 3.20 & 4.19 \\
		\textbf{EarlyF/A} & 1.46 & 9.57 & 0.76 & 2.85 & 15.93 & 0.66 & \underline{4.42} \\
		EarlyF/--- & 1.44 & \underline{11.24} & 0.79 & 2.70 & 15.89 & 2.62 & 3.83 \\
		MidF/B & 1.46 & 9.41 & 0.73 & 2.94 & 13.39 & -0.99 & 4.05 \\
		MidF/A & \underline{1.49} & 9.21 & 0.68 & 2.91 & 16.01 & -1.08 & 4.05 \\
		MidF/--- & 1.47 & 8.48 & 0.53 & 3.00 & 15.68 & 0.85 & 3.60 \\
		LateF/B & 1.43 & \textbf{11.67} & 0.76 & 2.64 & \textbf{18.51} & -0.62 & 4.20 \\
		LateF/A & 1.46 & 10.12 & 0.57 & 2.82 & 14.90 & 2.39 & 4.20 \\
		LateF/--- & 1.44 & 9.56 & 0.41 & 2.80 & \underline{17.74} & 1.19 & 3.91 \\
		\hline
		EarlyF/A+ & 1.49 & 18.78 & 4.44 & 2.38 & 19.96 & 14.95 & 3.52
	\end{tabular}
	
	\vspace{2mm}
	\caption{Experiment results for all models as in Table~\ref{tab:unseen_res_sn_0}, but with \textbf{echo} training target \textit{\textbf{OutD}}: $\overline{D}_\ell(k) \! = \! D_\ell(k)$, and subsequent subtraction from the microphone signal. Best result per measure is marked in \textbf{bold} font, second best is \underline{underlined}. Additional result for best model LateF/A with separate subsequent noise reduction from~\cite{strake_SingleStage_ICASSP}, retrained on this work's data (LateF/A+).}
	\label{tab:unseen_res_d_0}
	\vspace{2mm}
	\setlength\tabcolsep{0pt} 
	\begin{tabular}{p{14mm}<{\centering} | p{9mm}<{\centering} p{12mm}<{\centering} p{12mm}<{\centering} p{12mm}<{\centering} | p{9mm}<{\centering} | p{9mm}<{\centering} | p{8mm}<{\centering}}
		\textbf{Model/} & \multicolumn{4}{c|}{full mixture} & $d(n)$ & $n(n)$ & $s(n)$ \\
		\textbf{Skip} & \textbf{PESQ} & \textbf{ERLE}$_{\text{BB}}$ & \textbf{dSNR}$_{\text{BB}}$ & \textbf{PESQ}$_{\text{BB}}$ & \textbf{ERLE} & \textbf{dSNR} & \textbf{PESQ} \\
		\hline
		Kalman & 1.15 & 3.49 & -0.94 & \textbf{4.64} & \textbf{18.36} & --- & \textbf{4.64} \\
		\hline
		EarlyF/B & 1.94 & 8.71 & 0.68 & 3.21 & 7.89 & 4.40 & 4.26 \\
		EarlyF/A & 1.86 & \textbf{9.75} & 0.47 & 2.66 & 7.81 & 2.02 & 4.45 \\
		EarlyF/--- & 1.93 & 6.56 & 0.72 & \underline{3.68} & 6.70 & 3.50 & 4.34 \\
		MidF/B & \underline{1.98} & 7.90 & 0.69 & 3.28 & 8.50 & 4.51 & 4.02 \\
		MidF/A & 1.89 & 7.17 & 0.87 & 3.14 & 14.49 & 4.27 & 4.46 \\
		MidF/--- & 1.92 & 6.73 & 0.74 & 3.57 & 7.48 & 1.81 & 4.32 \\
		LateF/B & 1.88 & \underline{9.17} & 0.50 & 3.01 & 13.66 & 3.63 & 4.02 \\
		\textbf{LateF/A} & 1.86 & 9.06 & 0.69 & 3.17 & \underline{16.65} & 0.36 & \underline{4.50} \\
		LateF/--- & \textbf{2.10} & 5.85 & 0.36 & 3.59 & 2.87 & 2.06 & 4.47 \\
		\hline
		LateF/A+ & 1.58 & 18.42 & 5.29 & 2.66 & 20.65 & 21.23 & 3.55
	\end{tabular}
\end{table}

\vspace{-1.3mm}
Experiment results for all combinations of our proposed variants are shown in Tables~\ref{tab:unseen_res_s_0} to~\ref{tab:unseen_res_d_0} using three measures to rate the perfomance in different categories: the recently updated wide\-band PESQ MOS LQO~\cite{ITU_P862.2, ITU_P862.2_Corr} for speech quality, SNR improvement as deltaSNR (dSNR) in [dB] for noise reduction, and echo return loss enhancement (ERLE) as $ERLE(n) = 10 \cdot \log \left(d^2(n) / (d(n)-\hat{d}(n))^2 \right)$
for echo suppression. The final ERLE is computed as in \cite{jung_shadowwidebandfdaf} using first order IIR smoothing with factor 0.9996 on each of the samplewise components $d(n)$ and $\hat{d}(n)$, and is averaged over whole files.

Each table is split into two major parts: the three right-most columns provide the network performance when the input files consist either of echo only ($d(n)$, rated with ERLE), or of near-end noise ($n(n)$, rated with dSNR) or near-end speech only ($s(n)$, rated with PESQ). 
These results allow deep insights into each network model: How does it handle echo or near-end noise if no other signals are present? And most importantly: Can the model \emph{'simply'} pass through clean near-end speech?

The four center columns provide results for the normal previously described test sets, i.e., \emph{full mixture} input signals. Here, the PESQ MOS for the full output signal is evaluated, and 
the so-called black box approach according to ITU-T Recommendation~P.1110~\cite[sec. 8]{ITU_P1110} and \cite{fingscheidt_signalseparation, fingscheidt_blackbox, steinert_instrumentaldistortionassessment} is used to obtain the processed \textit{components} of the enhanced signal $e(n) = \tilde{d}(n) + \tilde{n}(n) + \tilde{s}(n)$, thereby allowing to compute ERLE, dSNR, and PESQ on the separated processed components $\tilde{d}(n)$, $\tilde{n}(n)$, and $\tilde{s}(n)$, respectively. These measures are marked with the index \emph{BB} (for black box). 

To allow for a better rating of the results, we additionally provide the performance of a traditional AEC algorithm, the well-known variationally diagonalized state-space frequency domain adaptive Kalman filter including its residual echo suppression postfilter \cite{enzner_vary_fdaf, MalikBenesty_VDMS-FDAF-AEC, jung_elshamy_SAEC, Franzen_Book_ICC}, as reference point. 

Table~\ref{tab:unseen_res_s_0} shows the results for all models with \textbf{clean speech} training target \textit{\textbf{OutE}}: $\overline{E}_\ell(k) \! = \! S_\ell(k)$. 
The models without skip connections achieve higher echo and noise suppression with up to $21.33$\,dB~ERLE and $33.62$\,dB~dSNR when only the respective component is present at the microphone. A clear preference for the encoder fusion position cannot be seen for this target choice, but the early fusion model \emph{EarlyF/---} shows best overall trade-off results; note that Zhang et~al.~\cite{wang_NN_AEC_18, wang_NN_AEC_19} also perform early fusion with clean speech targets. 
However, the strong suppression performance of clean speech targets comes at a price: With the PESQ values not going above $3.65$\,MOS, none of the models is able to pass through clean speech. This can also be seen in the full mixture results, especially when compared with the perfect near-end speech component score $\text{PESQ}_{\text{BB}}$ of the Kalman filter reference.

With the \textbf{noisy speech} target choice \textit{\textbf{OutE}}: $\overline{E}_\ell(k) \! = \! S_\ell(k)  + \! N_\ell(k)$ in Table~\ref{tab:unseen_res_sn_0}, the PESQ scores are slightly increased, whereas the echo suppression performance is slightly decreased, when only the respective components are present at the microphone. It can be seen that skip connections are extremely helpful to pass through clean speech, and---considering the near-end speech quality---the best overall trade-off design for these targets is the early fusion model \emph{EarlyF/A}. However, the PESQ scores on the full mixture remain comparable to those in Table~\ref{tab:unseen_res_s_0}. The diversity of the results again shows how important the design choices are in order to find a good tradeoff between suppression performance and near-end speech quality.

Finally, the results for our newly proposed \textbf{echo} training target \textit{\textbf{OutD}}: $\overline{D}_\ell(k) \! = \! D_\ell(k)$, and subsequent subtraction from the microphone signal, are displayed in Table~\ref{tab:unseen_res_d_0}.
The later fusion positions proof highly beneficial and lead to the best model \emph{LateF/A} for these targets. In contrast to the previous tables, this model does not only achieve a high echo suppression but maintains the best near-end speech quality at the same time. 
While this specific model also outperforms the best trade-off model from Table~\ref{tab:unseen_res_sn_0}, \textit{the full mixture PESQ scores (leftmost column) for all models are clearly above all other target choices}. 

For the two best trade-off models in Tables~\ref{tab:unseen_res_sn_0} and~\ref{tab:unseen_res_d_0}, we considered to perform a subsequent separate noise reduction~\cite{strake_SingleStage_ICASSP} on output signal $e(n)$ as postprocessor after AEC, trained on this work's data (symbol\,+). 
The results are displayed in the bottom lines of the tables. As to be expected, they show improved noise and residual echo suppression, but interestingly reveal a degradation of near-end speech again---whereas only our proposed \emph{LateF/A}~DNN echo target AEC was able to maintain the near-end speech quality.

\vspace{-2.1mm}
\section{Conclusions}
\label{sec:conc}
\vspace{-1.7mm}
We presented a deeper investigation of acoustic echo cancellation with fully convolutional neural networks. Along with a newly proposed network structure in the form of an echo estimator that delivers a significantly improved near-end speech quality over existing approaches (model: LateF/A DNN, echo target, Table 3), we revealed trade-offs between different performance aspects in terms of echo suppression, noise reduction, and near-end speech quality, thereby giving the reader guidance on crucial design choices for the not yet widely explored field of DNN~AEC.

\vspace{-2mm}
\bibliographystyle{IEEEbib}
\bibliography{biblio_franzen}

\end{document}